# r-Process elements from magnetorotational hypernovae

D. Yong[1,2*], C. Kobayashi[3,2], G. S. Da Costa[1,2], M. S. Bessell[1], A. Chiti[4], A. Frebel[4], K. Lind[5], A. D. Mackey[1,2], T. Nordlander[1,2], M. Asplund[6], A. R. Casey[7,2], A. F. Marino[8], S. J. Murphy[9,1] & B. P. Schmidt[1]

[1]Research School of Astronomy & Astrophysics, Australian National University, Canberra, ACT 2611, Australia

[2]ARC Centre of Excellence for All Sky Astrophysics in 3 Dimensions (ASTRO 3D), Australia

[3]Centre for Astrophysics Research, Department of Physics, Astronomy and Mathematics, University of Hertfordshire, Hatfield, AL10 9AB, UK

[4]Department of Physics and Kavli Institute for Astrophysics and Space Research, Massachusetts Institute of Technology, Cambridge, MA 02139, USA

[5]Department of Astronomy, Stockholm University, AlbaNova University Center, 106 91 Stockholm, Sweden

[6]Max Planck Institute for Astrophysics, Karl-Schwarzschild-Str. 1, D-85741 Garching, Germany

[7]School of Physics and Astronomy, Monash University, VIC 3800, Australia

[8]Istituto Nazionale di Astrofisica - Osservatorio Astronomico di Arcetri, Largo Enrico Fermi, 5, 50125, Firenze, Italy

[9]School of Science, The University of New South Wales, Canberra, ACT 2600, Australia

**Neutron-star mergers were recently confirmed as sites of rapid-neutron-capture (r-process) nucleosynthesis[1–3]. However, in Galactic chemical evolution models, neutron-star mergers alone cannot reproduce the observed element abundance patterns of extremely metal-poor stars, which indicates the existence of other sites of r-process nucleosynthesis[4–6]. These sites may be investigated by studying the element abundance patterns of chemically primitive stars in the halo of the Milky Way, because these objects retain the nucleosynthetic signatures of the earliest**

**generation of stars**[7–13]. **Here we report the element abundance pattern of the extremely metal-poor star SMSS J200322.54−114203.3. We observe a large enhancement in r-process elements, with very low overall metallicity. The element abundance pattern is well matched by the yields of a single 25-solar-mass magnetorotational hypernova. Such a hypernova could produce not only the r-process elements, but also light elements during stellar evolution, and iron-peak elements during explosive nuclear burning. Hypernovae are often associated with long-duration γ-ray bursts in the nearby Universe**[8]. **This connection indicates that similar explosions of fast-spinning strongly magnetized stars occurred during the earliest epochs of star formation in our Galaxy.**

The SkyMapper telescope has surveyed the southern sky[14] and identified thousands of chemically primitive stars[15] including the red giant star SMSS J200322.54-114203.3 (hereafter SMSS 2003-1142). The metallicity of this star is [Fe/H] = –3.5 based on the analysis of the high-resolution spectrum described below and in the Methods. (Here [A/B] = $\log_{10}(N_A/N_B)_{star} - \log_{10}(N_A/N_B)_\odot$, where $N_A/N_B$ is the number ratio of atoms of elements A and B, and the subscript ☉ refers to the solar value). Given the very small amount of heavy elements, "metals", present in this object (the iron-to-hydrogen ratio is 3,000 times lower than that for the Sun), we postulate that all the elements in SMSS 2003-1142, from carbon to the r-process element uranium, were likely produced by a single zero-metallicity progenitor star. Of particular interest is the fact that SMSS 2003-1142 is one of a small number of extremely metal-poor stars that exhibit large enhancements in the abundances of the r-process elements.

We obtained high-resolution spectroscopic observations of SMSS 2003-1142 in order to conduct a detailed chemical abundance analysis and infer the properties of its progenitor (see Methods). We find, in particular, a high nitrogen to iron abundance ratio, [N/Fe] = +1.07 relative to the carbon abundance [C/Fe] < 0.07 (with minimal correction for evolutionary mixing; see Methods), strongly supporting a rapidly rotating progenitor star[16]. Furthermore, the high zinc abundance, [Zn/Fe] = +0.72, can only be explained as originating from supernovae with large explosion energy[8]. For all other elements between carbon and zinc, the abundance pattern of SMSS 2003-1142 lies within the distribution of abundance patterns exhibited by other extremely metal-poor stars (EMP; [Fe/H] < –3) and is likely due to a single enrichment event from a zero-metallicity progenitor.

We also measured the relative abundances (or upper limits) for 28 neutron-capture elements from strontium (Z = 38) to uranium (Z = 92). For all neutron-capture elements, the abundance ratios normalised to iron, [X/Fe], are higher than the solar ratio. Among the known stars with large r-process element abundance enhancements, i.e., those with [Eu/Fe] > +1.0, SMSS 2003-1142 is the most iron-poor object by about factor of two. We plot the r-process element abundances in Figure 1 and compare the data to the scaled solar r-process distribution. For the elements from barium (Z = 56) to thulium (Z = 69), SMSS 2003-1142 exhibits the scaled solar r-process pattern that is also characteristic of r-process enhanced stars[17]. For lighter elements (Z < 56) and heavier elements (Z > 69), SMSS 2003-1142 exhibits lower and higher average abundances, respectively, when compared to the scaled solar r-process distribution. Compared to other known r-process enhanced stars, SMSS 2003-1142 has the highest [X/Fe] abundance ratios for zinc, barium, europium, and thorium, further highlighting the unusual nature of this star[18] (see Methods).

In Figure 2, we compare the abundance pattern of SMSS 2003-1142 with nucleosynthesis yields of a magnetorotational hypernova from a zero-metallicity 25 M⊙ star to learn more about the enrichment source. The observed abundance pattern at Z < 31 can be well reproduced by energetic (>$10^{52}$ erg) core-collapse supernovae from massive stars (M > 25 M⊙) – hypernovae[8,13], while the pattern at Z > 37 requires the r-process, as in magnetorotational supernovae[4,19,20]. We thus propose magnetorotational hypernovae for the enrichment source, assuming that the hypernova-type event (i.e., an energetic supernova producing 0.017 M⊙ of iron) is associated with an ejection of neutron-rich matter (see Methods for more details). The mass of the neutron-rich ejecta, 0.00035 M⊙, is obtained by matching the observed [Eu/Fe] ratio with theoretical yields. This model can naturally reproduce many of the key features of the observed abundance pattern from carbon to uranium, including the normal [C/Fe], normal [α/Fe] (where α refers to the average of Mg, Si, and Ca), low [Mn/Fe], high [(Co, Zn)/Fe], enhancement of the first and second peaks of neutron-capture elements, low [Pb/Fe], and high [(Th, U)/Fe]. Around the second peak (barium to neodymium, inclusive) the model predicts lower abundance ratios than observed, but the agreement could be improved by fine-tuning the model parameters (see Methods). The observed enhancements of nitrogen and sodium are not produced because of the lack of stellar rotation in the pre-supernova model. Scandium and titanium are known to be underproduced in one-dimensional supernova models (and are usually excluded from EMP modelling)[12]. The low [Mn/Fe] ratio provides a constraint on the material from Type Ia supernovae[21]; for SMSS 2003-1142 there is no evidence for any Type Ia supernovae contribution, and the prediction of our preferred model is consistent with the observed ratio. While the abundance pattern from carbon to iron could be reproduced by normal-energy supernova models, the high [X/Fe] ratios for cobalt and zinc strongly favor a high explosion energy as in hypernovae.

Neutron-star mergers were first posited some 40 years ago as a site for r-process nucleosynthesis[22] which was confirmed with recent direct observations of astronomical transient 'kilonova' (AT2017gfo)[2] and a short gamma-ray burst[23] following the detection of the gravitational wave event GW170817. Hence, Figure 2 includes a neutron-star merger as an alternative model[24]. Consequently, we assume that the neutron-star merger occurs in an interstellar medium that has already been enriched by core-collapse supernovae up to [Fe/H] ~ –3.5 (see Methods for the details). In this model[18] the metallicity of [Fe/H] ~ –3.5 is reached some 60 Myr after the onset of galaxy formation. While the general abundance pattern of SMSS 2003-1142 can be matched, this model fails to reproduce some of the key features. Namely, the predicted [α/Fe] ratio is ~0.35 dex higher than the observed value. Also, the [(Co, Zn)/Fe] ratios are ~ 0.6 dex lower than for our preferred supernova model underscoring the importance of a hypernova contribution. Further, the predicted thorium and uranium abundances for the neutron-star merger model are inconsistent with the observations. On the other hand, the predicted abundances around the second peak (barium to neodymium, inclusive) happen to be higher than those for the magnetorotational supernova yields[20]. However, as discussed in the Methods, the mismatches of the magnetorotational hypernova model could be solved by nuclear fissions and/or future self-consistent simulations, and we emphasize that no special significance should be assigned to this difference.

We note also that the ~60 Myr Fe-enrichment timescale in the alternative model is much shorter than that expected for neutron-star mergers so that, at this epoch, the rate of neutron-star mergers is extremely low[18]. We regard this as further evidence that the neutron-star merger hypothesis is inferior to our magnetorotational hypernova scenario. In addition, not only does our preferred model provide a superior fit to the data, it also involves only a single enrichment event, in contrast to the neutron star merger scenario which requires multiple generations of star formation. Given the low metallicity of SMSS 2003-1142, a neutron star merger origin would require new constraints on the formation, evolution, and merger timescale for such objects, as the merger event must have occurred very shortly after the onset of galaxy formation. While it is possible that more than one star contributed to the enrichment of SMSS 2003-1142, it is not required to explain the data. In summary, our analysis of SMSS 2003-1142 reveals evidence for r-process nucleosynthesis from magnetorotational hypernovae that may have occurred before the first neutron-star - neutron-star mergers in the earliest stages of galaxy formation.

In Galactic chemical evolution models, the contribution from hypernovae is very important for explaining the behavior of the abundance patterns of elements such as zinc and cobalt[21]. In regard to the r-process elements, Galactic chemical evolution models cannot reproduce the observed r-process element abundance patterns using neutron star mergers alone, and other r-process sites such as magnetorotational supernovae have been supported[5,18]. However, the connection between hypernovae and magnetorotational supernovae, namely, the explosion mechanism, is uncertain. Numerical simulations show that neutrino-driven convection[25,26] can result in the explosion of stars with $M < 25$ $M_\odot$ leaving behind a neutron star. For more massive stars with $M > 25$ $M_\odot$ that leave behind a black hole remnant, it is not yet known how such objects explode. However, in the nearby Universe the explosions of such massive stars have been observed as broad-lined Type Ic supernovae and are often associated with long duration gamma-ray bursts, which implies that stellar rotation and/or magnetic fields produce a jet that ejects iron-rich matter from near the central black hole. In fact, collapsars[27] and magnetars[28] have been proposed as a model of the central engine of long duration gamma-ray bursts. The magnetorotational hypernova model we propose here has a similar mechanism. This particular class of stellar explosions can be more important in the early Universe because in primordial stars the lack of stellar winds results in substantially less mass loss. The discovery and analysis of SMSS 2003-1142 will hopefully stimulate the community to further study magnetorotational hypernovae.

**Table 1: Chemical abundances of SMSS 2003-1142.**

(The Table is located at the end of this document)

**Figure Legends**

**Figure 1: R-process element abundance pattern of SMSS 2003-1142. a,** The abundance pattern for SMSS 2003-1142 compared to the scaled solar r-process distribution normalized to the europium abundance. **b,** The abundance differences (SMSS 2003-1142 – solar r-process). Arrows denote upper limits. The error bars are 1σ estimates of the uncertainties in our measurements as described in the Methods.

**Figure 2: Comparison of models and data**. **a,** Element abundance pattern for SMSS 2003-1142 compared to the 25 $M_\odot$ magnetorotational hypernova (MRHN) and the neutron star merger (NSM) models. **b**, Differences (SMSS 2003-1142 – models) excluding abundance limits as well as the elements nitrogen, sodium, scandium, and titanium (see Methods). The RMS values are 0.44 and 0.47

dex for the MRHN and NSM models, respectively. For our preferred set of elements (Z < 56 and Z > 60, see Methods), the RMS values are 0.34 and 0.50 dex for the MRHN and NSM models, respectively. The error bars are estimates of the 1σ uncertainties in our measurements as described in the Methods.

(Figures are located at the end of this document)

**Methods**

**Observational data:** SMSS 2003-1142 was identified as a candidate extremely metal-poor (EMP; [Fe/H] < –3) star from photometry during the course of the SkyMapper search for EMP stars[15]. This star was subsequently observed with the WiFeS integral field spectrograph[29] on the Australian National University's 2.3m telescope at Siding Spring Observatory using the B3000 and R3000 gratings to provide a resolving power of R = 3,000. Comparison of the flux-calibrated spectra against a grid of model fluxes[30] indicated a metallicity of [Fe/H] = –3.75. SMSS 2003-1142 was therefore sufficiently metal-poor to be included in our list for follow-up study at high spectral resolution.

High-resolution spectroscopic observations of SMSS 2003-1142 were obtained using the Magellan Inamori Kyocera Echelle (MIKE) spectrograph[31] at the 6.5m Magellan telescope on 2017 October 9. The exposure time was 720 seconds and wavelength coverage from 3,400 Å to 9,000 Å was obtained. The observations were performed using a 1.0-arcsec slit (with 2x2 CCD binning) that provided a resolving power of R = 28,000 in the blue arm and R = 22,000 in the red arm. The spectra were reduced using the CARPY pipeline[32]. The signal-to-noise (S/N) ratio of the reduced spectrum was 86 per 0.045 Å pixel near 4,500 Å.

From our analysis of the MIKE spectrum, we confirmed the low metallicity nature of this object, finding [Fe/H] = –3.5, and recognised that the europium abundance was unusually high, [Eu/Fe] = +1.7 dex. In addition to iron and europium, we measured the relative abundances for 25 chemical elements and found enhancements in the [X/Fe] abundance ratios for 10 neutron-capture elements from strontium (Z = 38) to erbium (Z = 68). In order to conduct a more detailed chemical abundance analysis, we required additional observations extending to bluer wavelengths at higher spectral resolution and S/N.

We were awarded Director's Discretionary Time on the European Southern Observatory's (ESO) Very Large Telescope (VLT) using the Ultraviolet and Visual Echelle Spectrograph (UVES[33]) (ESO proposal 2103.D-5062(A)). Six sets of 1,500 second exposures were obtained on 2019 September 06. We selected the 390+580 setting which provided wavelength coverage from 3,300 Å to 4,500 Å in the blue arm and 4,800 Å to 6,800 Å in the red arm. We used the 1.0-arcsec slit in the blue (2x2 CCD binning) and the 0.3-arcsec slit in the red (1x1 CCD binning) that provided spectral resolutions of R = 40,000 and R = 110,000 in the blue and red arms, respectively. The data were reduced using the ESO Reflex environment and the UVES pipeline version 5.10.4. In the co-added spectra, the S/N near 3,400 Å was 100 per re-binned 0.027 Å pixel. Our analysis focused on the spectra from the blue arm.

**Stellar parameters and abundance determination:** Stellar parameters were obtained using our previous methods[34,35]. Briefly, the effective temperature was determined from fitting model atmosphere fluxes to the spectrophotometric observations from the Australian National University's 2.3m telescope. The surface gravity was adopted from isochrones assuming the effective temperature and an age of 10 Gyr. Dwarf/giant discrimination is obtained from the WiFeS spectrophotometric observations and confirmation of the giant evolutionary state is provided by *Gaia* EDR3[36] where the parallax indicates an absolute magnitude in G of –0.4. We report an effective temperature of $T_{\mathrm{eff}}$ = 5,175 K, a surface gravity of log $g$ = 2.44 (cgs), a microturbulent velocity of 1.9 km/s, and a metallicity of [Fe/H] = –3.5.

Element abundances were obtained from equivalent width measurements or spectrum synthesis[37], using the stellar line analysis program MOOG[38,39] and one-dimensional local thermodynamic equilibrium (LTE) model atmospheres[40]. Extended Data Figure 1 illustrates an example spectrum synthesis fit to Zn and Eu lines. Uncertainties in the stellar parameters were estimated to be effective temperature ($T_{eff}$) ± 100K, surface gravity (log $g$) ± 0.3, microturbulent velocity ± 0.3 km/s, and metallicity ± 0.1 dex, and element abundance uncertainties were obtained using standard procedures[37] where the uncertainties in the abundances were added in quadrature.

The [X/Fe] ratios were computed adopting the following approach. For neutral species (e.g., NaI, MgI, etc.), we used the iron abundance as determined from the neutral iron lines such that the ratios are [NaI/FeI], [MgI/FeI], etc. For the singly ionized species (e.g., TiII, SrII), we used the iron abundance from the singly ionized iron lines, e.g., [TiII/FeII], [SrII/FeII].

Non-LTE abundance corrections have been computed for Na, Mg, Al, Si, Ca, Mn, Fe, and Ba. When using all elements (excluding nitrogen, sodium, scandium, titanium, and upper limits) and including those non-LTE abundance corrections, the RMS values for the MRHN and NSM models are 0.44 and 0.46 dex, respectively. (For comparison, the LTE RMS values are 0.44 and 0.47 dex for the MRHN and NSM models, respectively). Therefore, the MRHN model is still preferred over the NSM model even when taking into account non-LTE abundance corrections where available.

The evolutionary correction for the C abundance[41] is only 0.01 dex. Therefore, we can essentially ignore any evolutionary mixing effects that potentially could decrease the C and raise the N surface abundances.

In Extended Data Figure 2, we compare the relative abundances for C, N, Zn, Ba, Eu, and Th for a subset of highly r-process enhanced objects as well as for other metal-poor stars[18]. SMSS 2003-1142 has the highest [(Zn, Ba, Eu, Th)/Fe] ratios, and high [N/Fe], when compared to these stars illustrating the unique nature of this object. We also note that among the r-process enhanced metal-poor stars, some exhibit detectable amounts of the radioactive elements thorium and/or uranium[17]. Comparison of those abundances with the scaled solar r-process element pattern then enables nucleo-chronometric age dating[17]. For a subset of those objects, however, the thorium and/or uranium abundances are higher than the scaled solar values, which would imply a negative age. These stars are referred to as actinide boost stars. We note that while SMSS 2003-1142 is the most iron-poor of the actinide boost stars, it differs from other such objects in the [Th/Fe] vs. [Zn/Fe] plane; specifically SMSS 2003-1142 has higher ratios for both thorium and zinc by 0.2 dex. Moreover, when compared to stable r-process element abundances, the age for SMSS 2003-1142 inferred from the radioactive decay of Th and U is extremely uncertain, with estimates ranging from –11 to +11 Gyr.

**Kinematics:** This object is also known as Gaia DR2 4190620966764303488. Based on independent studies, this star has typical kinematics for the Milky Way halo population albeit with a retrograde orbit[42,43]. The radial velocities from the MIKE spectrum, UVES spectrum and Gaia DR2 are in good agreement; –50.7, –51.2, and –52.2 km/s, respectively.

**Theoretical Modelling:** In Figure 2, we show the nucleosynthesis yields from a zero-metallicity 25 $M_\odot$ magnetorotational hypernova (MRHN) model. There is no direct observation of such a supernova, and no successful explosion simulation either[44,45,46]. Consequently, there are no self-consistent nucleosynthesis yields available in the literature. Here we propose a hypernova-like core-collapse explosion associated with the ejection of neutron-rich matter. The ejection of neutron-rich matter from a massive star driven by stellar rotation and magnetic fields has been modelled in numerical simulations by a few research groups. We take the nucleosynthesis yields up to uranium (Z = 1-92) from a post-processing nucleosynthesis calculation[20] (the model B11β1.00) of a 2D special relativistic magnetohydrodynamic (MHD) simulation[47] for an iron core from a rotating 25 $M_\odot$ star with solar metallicity (which is the only metallicity available; however, the nucleosynthesis does not depend on the initial metallicity, although it affects the mass of the iron core). Because of the simulation setting, it is unknown how much envelope outside of the iron core is ejected into the interstellar medium and how much material falls back onto the central black hole during the explosion. We thus produced hypothetical nucleosynthetic yields for the whole star including the envelope. We choose a 25 $M_\odot$ hypernova model with the explosion energy of $10^{52}$ erg, no rotation, and zero metallicity from an existing set of supernova/hypernova models that are able to explain the observed abundances in EMP stars[48]. The model includes the nucleosynthesis yields up to germanium (Z = 32). It produces 0.017 $M_\odot$ of iron, which is 12 times larger than that in the original simulation of a magnetorotational supernova (MRSN), and is more luminous. Then, the relative contribution (i.e., the mass) of the neutron-rich ejecta in our MRHN model is determined by matching the observed [Eu/Fe] ratio; it is 0.00035 $M_\odot$, only 1.6% of the ejected matter in the MRSN simulation. Finally, we assume that our MRHN exploded into a pristine interstellar medium from which this EMP star was born; the ejecta are diluted into 3 x $10^4$ $M_\odot$ of primordial gas (hydrogen, helium, and trace amounts of lithium), to generate a metallicity of [Fe/H] = –3.5.

Does the progenitor star have to be 25 $M_\odot$? No. It is possible to reproduce the observed abundance pattern with a 40 $M_\odot$ hypernova (3 x $10^{52}$ erg), which ejects 0.33 $M_\odot$ of iron, associated with the ejection of 0.007 $M_\odot$ of neutron-rich matter, and diluted into 6 x $10^5$ $M_\odot$ of primordial gas. A 15 $M_\odot$ supernova ($10^{51}$ erg) could also reproduce the observed [Mg/Fe] but would result in lower [(Ca, Co, Zn)/Fe] ratios than observed. The observed [Ni/Fe] abundance favours hypernova models over supernova models since the jet explosion can eject Ni that formed near the black hole. Therefore, we conclude that the enrichment source is a 25-40 $M_\odot$ zero-metallicity star.

What is a MRHN? -- It is a jet explosion triggered by magnetic fields and core rotation, associated with ejection of neutron-rich matter, as shown in MHD simulations[4,20,49]. However, the ejected iron mass in our model is larger than in the simulations. Hence, our MRHN is more luminous, a few magnitudes brighter, than the theoretical MRSN (and several magnitudes brighter than a kilonova) but can be fainter or brighter than the supernovae observed in the nearby Universe. MRHN may also be related to long duration gamma-ray bursts and/or super-luminous supernovae. The central engine may be similar to those in collapsar[27] or magnetar[28] models. Currently, neither yields nor base explosion simulations are available, as both of which require three-dimensional hydrodynamical simulations that include general relativity and magnetic fields. The discovery and analysis of SMSS 2003-1142 will hopefully stimulate the community to further study MRHN.

Production of neutron-capture elements by the explosion of a massive rotating star has also been suggested as one of the possible explanations of the abundance distribution in the r-process rich star RAVE J183013.5-455510[50]. That object, however, has considerably different abundances when compared to SMSS 2003-1142, e.g., SMSS 2003-1142 has much lower C and higher Eu and Th abundances. As for alternative possibilities, it is not possible to explain the Th abundances with a spin star model[51]. It may be possible that SMSS 2003-1142 was enriched by multiple Population III supernovae; this possibility has been discussed in relation to the multiplicity of the first stars[52]. However, with multiple enrichment events it would be difficult to explain the low [Fe/H] of normal C abundance stars such as SMSS 2003-1142.

Figure 2 also shows an alternative model involving a neutron-star merger. Neutron stars form after supernova explosions of massive stars ($> 8M_\odot$)[8], and binary systems of two neutron stars are observed. Nevertheless, it takes a finite amount of time for two neutron stars to merge, the so-called delay-time. There are various binary population synthesis calculations that predict the distribution of the delay-times[53,54,55]. Therefore, it is likely that chemical enrichment has occurred prior to the first neutron-star mergers. Consequently, for the background interstellar medium, we take the elemental abundances from a Galactic chemical evolution model[18], where [Fe/H] reaches $\sim -3.5$ at t $\sim$ 60 Myr. This is before Type Ia supernovae start to occur. Asymptotic giant branch stars and electron-capture supernovae, both of which could produce some neutron-capture elements, do not contribute either at such an early time.

We then add the enrichment from a neutron-star merger to the background interstellar medium composition. We use post-processing nucleosynthesis yields from a 3D general relativistic simulation involving a merger of two 1.3$M_\odot$ neutron stars[24]. The ejecta mass of the simulation is 0.01 $M_\odot$, which is mixed into 3 x 10$^8$ $M_\odot$ of the slightly-enriched interstellar medium (with [Fe/H] ~ –3.5) in order to match the observed [Eu/Fe], prior to the formation of SMSS 2003-1142.

There are two crucial problems in this neutron-star merger scenario, namely the rate and the delay-time. The adopted Galactic chemical evolution (GCE) model self-consistently includes chemical enrichment from all stellar masses from 0.01 to 50 $M_\odot$, assuming the Kroupa initial mass function (hypernovae are not included). The star formation history is constrained to match the observed metallicity distribution function of stars in the solar neighborhood. In the model, the metallicity of the interstellar medium reaches [Fe/H] ~ –3.5 at 60 Myr after the onset of the star formation. At this time not many stars have formed, and thus the number of double neutron star systems is extremely small. Moreover, the timescale is too short compared with the typical delay-time of neutron-star mergers which is ~100 Myr[53,54,55], or longer (the delay-time distribution depends on the parameters of the binary population synthesis). In a similar model for the Galactic halo[18], chemical enrichment takes place quickly but inefficiently; although the number of double neutron star systems are slightly larger, the [Fe/H] reaches ~ –3.5 at t = 10 Myr, which makes it even harder to have a neutron star merger. Alternatively, it is possible to have interstellar matter that is locally less enriched compared with the average of the Galaxy at a given time. This inhomogeneous enrichment effect can be important for dwarf satellite galaxies, and in a GCE model for ultra-faint dwarf galaxies the [Fe/H] could stay below –3.5 at t = 300 Myr. However, this effect should be studied with more self-consistent hydrodynamical simulations, though the frequency of the enrichment from neutron star mergers at [Fe/H] ~ –3.5 is found to be extremely low in the simulations of a Milky Way-type galaxy[5,56].

We should also note that the detailed abundance pattern above germanium ($Z$ = 32) depends on the parameters in the r-process calculations, such as the strength of rotation and magnetic field for

MRSN[20,49], the mass and equation of state of neutron stars[24,57], the detailed modelling of neutrino transport[45] and nuclear fissions, and on nuclear reaction rates. We find in the MRSN model that the elements around $A \sim 90$ are overproduced, while the elements around $A \sim 140$ are underproduced, which may suggest a possible problem with fission modelling. The other effects can change the resultant distribution of the electron fraction ($Y_e$) in the simulations. Matter with $Y_e < 0.5$ is neutron-rich, and lower values of $Y_e$ tend to produce heavier elements. The high abundance of thorium and uranium are caused by matter with $Y_e < 0.1$ in the adopted MRSN simulation, but are not reproduced in the adopted NSM simulation. The relatively high abundances of the second (barium) peak elements are caused by the matter with $Y_e \sim 0.2$ in the NSM simulation. These elements are underproduced in the MRSN simulation, but could be increased with different parameters and/or in future self-consistent 3D+MHD+GR simulations. Alternatively, the barium abundance might be enhanced by stellar rotation[51], but this would result in even higher [(Sr,Y,Zr)/Fe] ratios than in the models presented in Figure 2, and hence is not a preferred solution.

Finally, following previous work[12], we note that when computing the RMS values provided in the caption for Figure 2, we excluded N and Na since the model does not include stellar rotation. We also exclude Sc and Ti since these elements are known to be underproduced in one-dimensional supernova models[12]. We also exclude the elements from Ba to Nd, inclusive, because of the uncertainties discussed above. Regardless of whether we use our preferred set of elements ($Z < 56$ and $Z > 60$ and excluding nitrogen, sodium, scandium, and titanium) or all elements (excluding nitrogen, sodium, scandium, and titanium), the RMS values favor the MRHN model over the NSM model.

**Methods references**

29. Dopita, M. et al. The Wide Field Spectrograph (WiFeS): performance and data reduction. *Astrophysics and Space Science* **327**, 245-257 (2010).

**Data Availability:** The data used in this study are available in the ESO archive (https://archive.eso.org/eso/eso_archive_main.html) under program ID 2103.D-5062(A).

**Code Availability:** The stellar line analysis program MOOG is available online at https://www.as.utexas.edu/~chris/moog.html. The stellar model atmospheres are available online at http://kurucz.harvard.edu/grids.html.

**Acknowledgements:** This paper includes data gathered with the 6.5-m Magellan Telescopes located at Las Campanas Observatory, Chile. Based on observations collected at the European Southern Observatory under ESO programme DDT 2103.D-5062(A). This research was supported by the Australian Research Council Centre of Excellence for All Sky Astrophysics in 3 Dimensions (ASTRO 3D), through project number CE170100013. C.K. acknowledges funding from the UK Science and

Technology Facility Council (STFC) through grant ST/M000958/1 & ST/ R000905/1, and the Stromlo Distinguished Visitor Program at the ANU. KL acknowledges funds from the European Research Council (ERC) under the European Union's Horizon 2020 research and innovation programme (Grant agreement No. 852977). AFM acknowledges support from the European Union's Horizon 2020 research and innovation programme under the Marie Sklodowska-Curie grant agreement No 797100. ARC acknowledges Australian Research Council grant DE190100656.

**Author Contributions:** GDC, MSB, MA, ADM, AFM, SJM, and TN were involved in the target selection and low-resolution spectroscopic observation campaigns. DY, GDC, AC, AF, and TN were involved in the high-resolution spectroscopic observations. KL and TN computed non-LTE corrections. The manuscript was written by DY, CK, and GDC with contributions from all authors..

**Author Information:** Reprints and permissions information is available at www.nature.com/reprints. The authors declare no competing interests. Correspondence and requests for materials should be addressed to DY (email: david.yong@anu.edu.au).

**Additional information**

None

**Extended Data**

**Extended Data Figure 1: Spectrum of SMSS 2003-1142.** Spectrum synthesis fit to the 4,810 Å Zn I line (**a**) and the 4,129 Å Eu II line (**b**). The observed spectra are shown as small circles, the best fitting synthetic spectrum as the solid black line, and the yellow region indicates ± 0.2 dex from the best fit.

**Extended Data Figure 2: Abundance ratios in halo stars.** Element to iron ratios, [X/Fe], as a function of metallicity, [Fe/H], based on literature data[20] for C, N, Zn, Ba, Eu, and Th. The lines are the GCE model predictions for the solar neighborhood[20]. SMSS 2003-1142 is shown as the large 5-point star. The locations of well-studied r-process rich stars (CS 22892-052, HD 122563, CS 29497-004, CS 31082-001, RAVE J183013.5-455510) are highlighted by large symbols.

**Table 1: Chemical abundances of SMSS 2003-1142.**

| Species | Z | A(X) | N$_{lines}$ | s.e.o.m | [X/Fe] | Total Error |
|---|---|---|---|---|---|---|
| C (CH) | 6 | <4.93 | ... | … | <0.07 | 0.30 |
| N (NH) | 7 | 5.33 | ... | … | 1.07 | 0.30 |
| NaI | 11 | 2.80 | 2 | 0.09 | 0.13 | 0.11 |
| MgI | 12 | 4.26 | 7 | 0.03 | 0.23 | 0.07 |
| AlI | 13 | 1.94 | 1 | … | -0.94 | 0.16 |
| SiI | 14 | 4.27 | 1 | … | 0.33 | 0.17 |
| CaI | 20 | 3.11 | 7 | 0.02 | 0.34 | 0.08 |
| ScII | 21 | 0.09 | 1 | … | 0.37 | 0.19 |
| TiI | 22 | 1.83 | 1 | … | 0.45 | 0.17 |
| TiII | 22 | 1.77 | 17 | 0.05 | 0.26 | 0.12 |
| CrI | 24 | 1.76 | 5 | 0.07 | -0.31 | 0.10 |
| MnI | 25 | 1.27 | 3 | 0.04 | -0.59 | 0.10 |
| FeI | 26 | 3.93 | 91 | 0.02 | -3.57 | 0.11 |
| FeII | 26 | 4.07 | 6 | 0.06 | -3.43 | 0.13 |
| CoI | 27 | 1.78 | 3 | 0.08 | 0.36 | 0.11 |
| NiI | 28 | 2.90 | 3 | 0.07 | 0.25 | 0.09 |
| CuI | 29 | <1.30 | 1 | ... | <0.68 | 0.16 |
| ZnI | 30 | 1.71 | 1 | ... | 0.72 | 0.16 |
| SrII | 38 | 0.20 | 2 | 0.02 | 0.76 | 0.17 |
| YII | 39 | -0.67 | 7 | 0.03 | 0.55 | 0.09 |
| ZrII | 40 | -0.05 | 12 | 0.02 | 0.80 | 0.07 |
| MoI | 42 | -0.48 | 1 | ... | 1.21 | 0.16 |
| RuI | 44 | -0.10 | 2 | 0.04 | 1.72 | 0.12 |
| RhII | 45 | <-0.78 | 1 | ... | <1.74 | 0.16 |
| PdI | 46 | <-0.72 | 1 | ... | <1.28 | 0.16 |
| AgI | 47 | <-1.71 | 1 | ... | <0.92 | 0.16 |
| BaII | 56 | -0.10 | 4 | 0.03 | 1.15 | 0.11 |
| LaII | 57 | -1.03 | 7 | 0.02 | 1.30 | 0.09 |
| CeII | 58 | -0.80 | 3 | 0.08 | 1.05 | 0.11 |
| PrII | 59 | -1.18 | 1 | ... | 1.53 | 0.17 |
| NdII | 60 | -0.56 | 8 | 0.03 | 1.45 | 0.09 |
| SmII | 62 | -0.79 | 4 | 0.03 | 1.68 | 0.10 |
| EuII | 63 | -1.21 | 4 | 0.02 | 1.70 | 0.10 |
| GdII | 64 | -0.65 | 6 | 0.03 | 1.71 | 0.09 |
| TbII | 65 | -1.32 | 1 | ... | 1.81 | 0.16 |
| DyII | 66 | -0.56 | 14 | 0.02 | 1.77 | 0.07 |
| HoII | 67 | -1.31 | 2 | 0.01 | 1.64 | 0.12 |
| ErII | 68 | -0.80 | 5 | 0.03 | 1.71 | 0.09 |
| TmII | 69 | -1.53 | 3 | 0.09 | 1.80 | 0.11 |
| YbII | 70 | -0.60 | 1 | ... | 1.99 | 0.16 |
| LuII | 71 | -0.94 | 1 | ... | 2.39 | 0.16 |
| HfII | 72 | -1.12 | 1 | ... | 1.46 | 0.16 |
| OsI | 76 | 0.00 | 1 | ... | 2.17 | 0.16 |
| PbI | 82 | <-0.10 | 1 | ... | <1.72 | 0.16 |
| ThII | 90 | -1.31 | 2 | 0.09 | 2.10 | 0.12 |
| UII | 92 | -1.91 | 1 | ... | 2.06 | 0.16 |

Figure 1

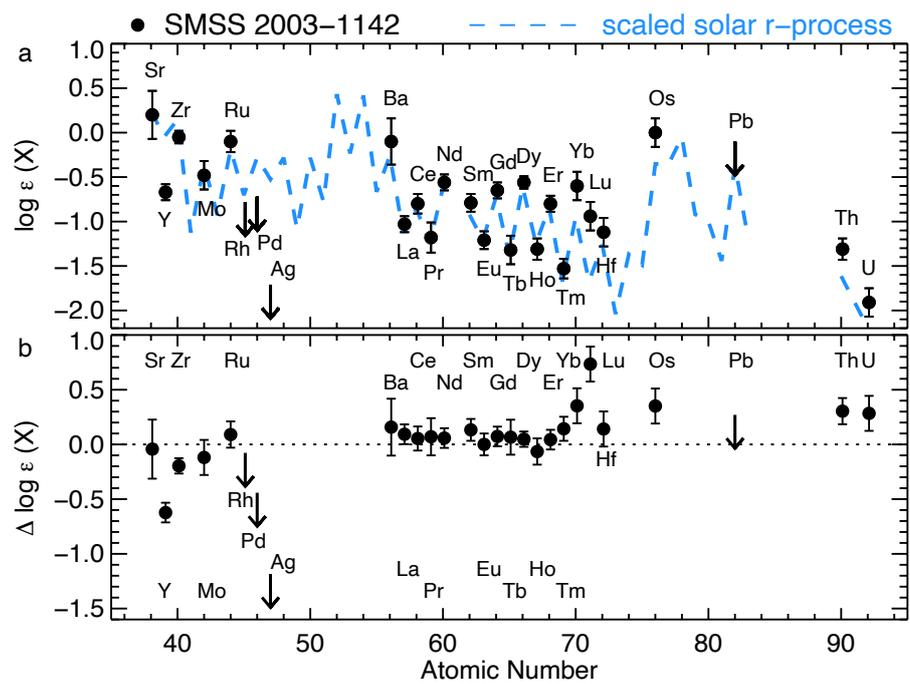

Figure 2

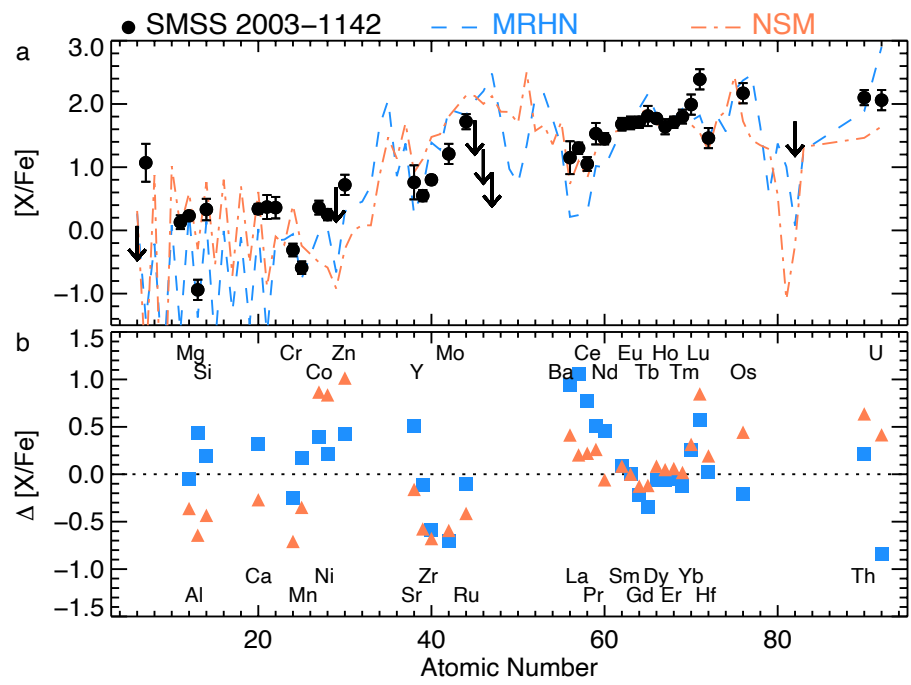

Extended Data Figure 1

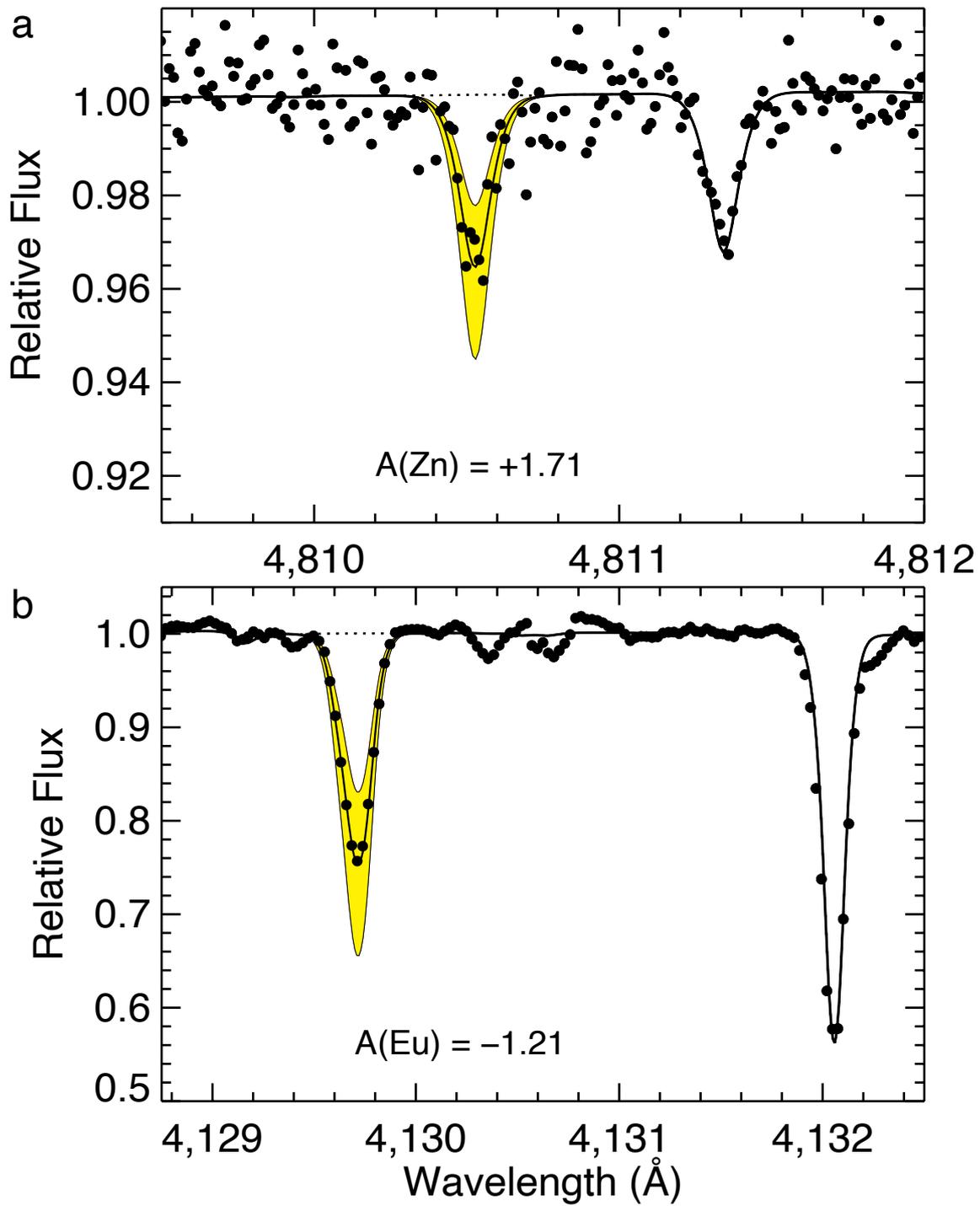

Extended Data Figure 2

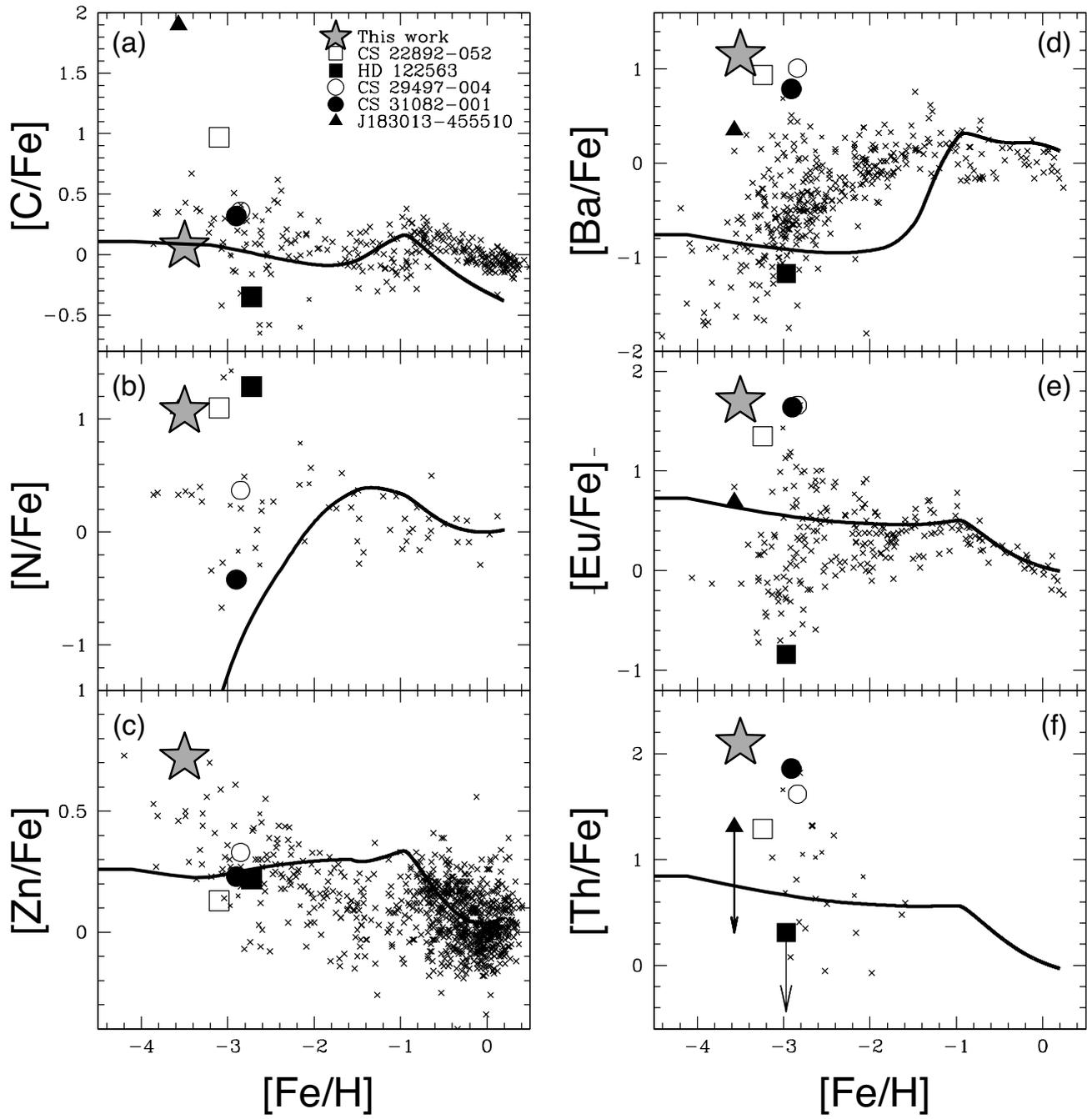